\documentclass{revtex4}
\def\[{\left\lbrack}
\def\]{\right\rbrack}

\def\({\left(}
\def\){\right)}
\newcommand{\be}{\begin{equation}}
\newcommand{\ee}{\end{equation}}
\newcommand{\ea}{\end{eqnarray}}
\newcommand{\ba}{\begin{eqnarray}}

\newcommand{\1}{{(1)}}

\newcommand{\vx}{{\vec{x}}}
\newcommand{\vy}{{\vec{y}}}

\newcommand{\vep}{{\varepsilon}}
\newcommand{\ep}{{\epsilon}}

\newcommand{\cl}{{\cal L}}
\newcommand{\cg}{{\cal G}}
\newcommand{\dirac}{{\delta(\vx - \vy)}}

\begin{document}

\title{Embedding Commutative and Noncommutative Theories in the Symplectic Framework}

\author{C. Neves, W. Oliveira}
\email{cneves, wilson@fisica.ufjf.br}
\affiliation{Departamento de F\'{\i}sica, Universidade Federal de Juiz de Fora,\\
36036-330, Juiz de Fora, MG, Brasil}
\author{D. C. Rodrigues and C. Wotzasek}
\email{clovis, cabral@if.ufrj.br}
\affiliation{Instituto de F\'{\i}sica, Universidade Federal do Rio de Janeiro,\\
21945-970, Rio de Janeiro, RJ, Brasil}

\begin{abstract}
\noindent
This paper is devoted to study gauge embedding of either commutative and noncommutative theories in the framework of the symplectic formalism\cite{f-j,b-w}.  We illustrate our ideas in the Proca model, the irrotational fluid model and the noncommutative self-dual model. In the process of this new path of embedding, the infinitesimal gauge generators of the gauge embedded theory  are easily and directly chosen. Among other advantages, this enables a greater control over the final Lagrangian and puts some light on the so called ''arbitrariness problem".

\end{abstract}

\maketitle

{\small Keywords: Noncommutativity, Symplectic, Gauge Embed.}

\setlength{\baselineskip} {20 pt}

\section{Introduction}

Three physical models, with very distinct properties, are here gauge embedded through the new method proposed in this paper. The first one, the Proca model,  to be discussed in Section III, is a classical example of theory without gauge symmetry due to a mass term. Using the Dirac nomenclature\cite{Dirac}, it is classified as a second class system. On the other hand, the irrotational fluid model, the second theory dealt with here (Section IV), cannot be in the same way classified, because it does not possess any constraints. Besides that, its Lagrangian has a potential term ( $1/\rho$, see (\ref{00010})) which could bring, at first sight, some difficulties to the method. The third and last model is the noncommutative self-dual model, discussed in Section V. This is a topologically massive second class theory with a Chern-Simons-like term. More important, the fields do not commute --- a feature that, just like the non-Abelian algebra, causes some trouble to other gauge embedding methods \cite{kpr, procabarcelos}. The noncommutative self-dual model, as the non-Abelian self-dual model, has been in the scope of many recent papers, its properties and dualities (under some limits) still need further investigations. An overview of the method is presented in the next section.

One of the greatest merits of this new method, which was inspired on \cite{jwc}, comes from its simple and direct way of {\it choosing} the infinitesimal gauge generators of the built gauge theory. This give us a freedom to choose the content of the embedded symmetry according to our necessities. This feature makes possible a greater control over the final Lagrangian. For example, with the BFT \cite{bft} method, noncommutative and non-Abelian theories are usually embedded into theories with infinite terms in the Hamiltonian and with infinitesimal gauge generators that cannot be expressed in closed form \cite{kpr, procabarcelos}. This can be avoided with the present method, because the infinitesimal gauge generators are not deduced from previous unclear choices, but, instead, are directly chosen.

Another related advantage is the possibility of doing a kind of ''general embedding", that is, instead of choosing the gauge generators at the beginning, one can leave some unfixed parameters with the aim of fixing them latter, when the final Lagrangian has being achieved. Although one can reach faster the final theory fixing such parameters as soon as possible, this path is more interesting in order to study the considered theory, and is helpful if the desired symmetry is unknown, but some aspect of the Lagrangian is wanted. This path of ''general embedding" was employed on each of the applications.

Lastly, we should mention that this approach to embedding is not depended on any undermined constraint structure and also works for unconstrained systems.  This is different from all the existent embedding techniques that use to convert\cite{bft,cw}, project \cite{wis} or reorder \cite{mitra} the existent second-class constraints into a first-class system. This technique on the other hand only deals with the symplectic structure of the theory so that the embedding structure independs on any pre-existent constrained structure. This is a new feature that will be explored in this paper.


\section{Overview of the method}

The proposed embedding method is grounded on the symplectic formalism \cite{f-j,b-w}. This formalism starts from a first order Lagrangian (density) written as
\be
	\label{LL}
	\cl (\xi, \dot \xi) = a_\alpha (\xi)\dot \xi^\alpha - V(\xi),
\ee
where the symplectic coordinates $\xi^\alpha$ are functions of the $n+1$-dimensional space-time, and the dependence on spatial derivatives is implicit.  To turn explicit their spatial dependence we use $\xi^\alpha (\vx)$. If the Lagrangian is not of first order, one can introduce the canonical momenta as new coordinates and linearize the Lagrangian (as we will see in next section). After that, we are able to compute the matrix
\be
	\label{CW1}
	f_{\alpha \beta}(\vx,\vy) = \frac{\delta a_\beta (\vy)}{\delta \xi^\alpha (\vx)} - \frac{\delta a_\alpha (\vx)}{\delta \xi^\beta (\vy)}.
\ee

	As stated by the symplectic formalism, the above matrix, called symplectic matrix, is degenerated if, and only if, there are some unknown constraints or there is gauge symmetry. The constraints in the symplectic formalism emerge from the calculation of
\be
	\int d^ny \; \nu^\alpha_k(\vx) \frac{\delta V(\vy)}{\xi^\alpha(\vx)},
	\label{vincintro}
\ee
where $(\nu^\alpha)_k$, for each $k$, is a zero-mode of the symplectic matrix.

	Some zero-modes may lead to non-null ({\it a priori}) expressions, which are the constraints. If at least one zero-mode does not generate a new constraint, that is, if (\ref{vincintro}) is null (or proportional to others known constraints) for some $k$, then the considered theory has gauge symmetry \footnote{Here it is being supposed a non-null potential $V$. A null potential is related to reparametrization invariance.}. In this case, the infinitesimal generators of the gauge symmetry are the components of the corresponding zero-mode \cite{mon}. Let the zero-mode $(\nu^\alpha)_{k_0}$ satisfy
\be
	\label{gaugeintro}
	\int d^ny \; \nu^\alpha_{k_0} (\vx) \frac{\delta V(\vy)}{\xi^\alpha(\vx)} = C^m(\vx) \Omega_m(\vx),
\ee
where the $C^m$'s are some functions of $\xi^\alpha$ and the $\Omega_m$'s are the known constraints. Then $(\nu^\alpha)_{k_0}$ inform us of the presence of gauge symmetry, whose infinitesimal generators are given by
\be
	\label{introLthetagen}
	\delta_{\vep_{k_0}} \xi^\alpha (\vx) = \int d^ny \; \vep_{k_0} (\vy) \nu^\alpha_{k_0} (\vy) \dirac,
\ee
with $\vep_{k_0}$ as the infinitesimal parameter. The integral and the delta function above are important because zero-modes some times have spatial derivatives.

	After this brief review of some properties of the symplectic formalism, lets see the key features of the embedding procedure. The first step is to insert new fields in the Lagrangian in such a way that, for a certain gauge (called unitary gauge), they vanish and we return to the original theory. These are called Wess-Zumino (WZ) fields \cite{f-s}. The most natural way of doing so, with just one WZ field, probably is
\be
	\label{introLtheta}
	\tilde \cl_\theta = \cl + \Psi \dot \theta - G,
\ee
where $\theta$ is the WZ field, $\Psi \equiv \Psi(\xi,\theta)$, $G \equiv G(\xi,\theta)$ and $G$   complies with $G(\theta=0)=0$. This structure is useful and it will be used in Section IV, but, depending on the structure of $\cl$, one can achieve better results with another WZ field, being responsible for the second order velocity term in the WZ sector after a Legendre transformation, i.e.,
\be 
	\label{ltg}
	\tilde \cl_{\theta,\gamma} = \cl + (\Psi + \gamma) \dot \theta - G - \frac k2 \gamma \gamma.
\ee
Therefore, the Euler-Lagrange equation for $\gamma$ is $\dot \theta = k \gamma$, and after $\gamma$ elimination the term $\dot \theta \dot \theta$ appears in the Lagrangian.

As we will see, $\tilde \cl_{\theta,\gamma}$ enables us to select gauge symmetries with temporal derivative on the infinitesimal parameter $\vep$, whereas $\tilde \cl_\theta$ can be used to theories without any constraints.

Let $f\equiv \left(f_{\alpha \beta}\right)$ be the symplectic matrix of $\cl$ as defined in (\ref{CW1}), then the symplectic matrix of $\tilde \cl_\theta$ and $\tilde \cl_{\theta,\gamma}$ are
\be
	\tilde f_\theta (\vx,\vy) = \pmatrix{ (f_{\alpha \beta}) & \frac{\delta \Psi(\vy)}{\delta \xi^\alpha(\vx)} \cr
	- \frac{\delta \Psi(\vx)}{\delta \xi^\beta(\vy)} & \Theta_{xy}},
\ee
\be
	\tilde f_{\theta, \gamma} (\vx,\vy) = \pmatrix{(f_{\alpha \beta}) & \frac{\delta \Psi(\vy)}{\delta \xi^\alpha(\vx)} & 0_\alpha \\[0.1in] \cr 
	- \frac{\delta \Psi(\vx)}{\delta \xi^\beta(\vy)} & \Theta_{xy} & -\dirac \\[0.1in] \cr 
	0_\beta & \dirac & 0}.
\ee
The symbol $\Theta_{xy}$ is defined as
\be
	\Theta_{xy} \equiv \frac{\delta \Psi(\vy)}{\delta \theta (\vx)} - \frac{\delta \Psi(\vx)}{\delta \theta (\vy)},
\ee
whereas $0_\alpha$ is a null column and $0_\beta$ is a null line.

	After $\tilde f_\theta$ is computed , the next step is already the selection of the gauge generators, i.e., the zero-mode that does not produce new constraints. Denoting this vector as $\tilde \nu$, the following conditions need to be fulfilled:
\be
	\label{mzcondftheta}
	\int d^nx \; \tilde \nu^\alpha (\vx)\tilde f^\theta_{\alpha \beta} (\vx,\vy) = 0,
\ee
\be
	\label{gaugecondftheta}
	\int d^ny \; \tilde \nu^\alpha (\vx) \frac {\delta \tilde V_\theta(\vy)}{\delta \tilde \xi^\alpha(\vx)} = 0,
\ee
where $\tilde V_\theta = V + G$ and $\tilde \xi^\alpha = (\xi^\alpha, \theta)$. Here the index $\alpha$ on symbols with tilde assumes an extra value in relation with the ones without tilde.

	The expression in (\ref{gaugecondftheta}), in general, needs only to be proportional to the known constraints, but at this stage no constraints have emerged from the symplectic algorithm (and none will be necessary while working with $\tilde \cl_\theta$).

	From equation (\ref{mzcondftheta}) the function $\Psi$ is found, and from (\ref{gaugecondftheta}) comes the solution for $G$. Therefore, with the described procedure, one can obtain the Lagrangian $\tilde \cl_\theta$, whose infinitesimal gauge generators are given by
\be
	\delta_{\vep} \tilde \xi^\alpha (\vx) = \int d^ny \; \vep (\vy) \tilde \nu^\alpha (\vy) \dirac.
\ee

Upon dealing with $\tilde \cl_{\theta, \gamma}$ the task is a little harder. We cannot hope that with such structure one can embed any possible physical system, but, as we will see in next sections, the structure of $\tilde \cl_{\theta, \gamma}$ is useful to handle important types of Lagrangian and provide us new and interesting answers to the embed problem. In next lines, the central ideas of the method, with no aim of being very general or rigorous, will be shown.

	After the computation of $\tilde f_{\theta,\gamma}$ one cannot immediately select the gauge generators, because the delta functions severely restrain our possibilities. The only possible form of the zero-mode is 
\be
	\label{introzmmod}
	\tilde \nu^\alpha(\vx) = ( \nu^\alpha(\vx), \; 0, \; b(\vx)).
\ee
The vector $(\nu^\alpha)$ is a zero-mode of $(f_{\alpha \beta})$, while $b$ is some function of $\xi^\alpha$ and $\theta$. To accomplish our purposes, $b$ will be selected as a non-null constant. Note that if the symplectic matrix $f$ does not possess zero-modes, then the same occurs to $\tilde f_{\theta, \gamma}$. Due to this, the employment of $\tilde \cl_{\theta, \gamma}$ is restricted to constrained systems (just like most of the gauge embedding procedures).

For $\tilde \nu$ be the zero-mode of $\tilde f_{\theta, \gamma}$ just one condition is necessary, namely,
\be
	\int d^nx \; \left ( \nu^\alpha \frac {\delta \Psi(\vy)}{\delta \xi^\alpha(\vx)} + b \dirac \right ) = 0,
\ee
which is our first equation for finding $\Psi$.

The vector $\tilde \nu$ will be useful to generate a constraint, providing us a more suitable symplectic matrix to our purposes. The constraint comes from 
\ba
	\int d^ny \; \tilde \nu^\alpha(\vx) \frac {\delta \tilde V_{\theta,\gamma}(\vy)}{\delta \tilde \xi^\alpha(\vx)} & = & \int d^ny \; \tilde \nu^\alpha(\vx) \frac \delta {\delta \tilde \xi^\alpha(\vx)} \left( V(\vy) + G(\vy) + \frac k2 \gamma(\vy)\gamma(\vy) \right ) \nonumber \\[0.1in]
	\label{introGnu}
	& = & \Omega(\vx) + \int d^ny \; \nu^\alpha (\vx) \frac {\delta G(\vy)}{\delta  \xi^\alpha(\vx)} + k b \gamma (\vx)  \\[0.1in]
	& = & \Omega  + G_\nu + kb \gamma \equiv \tilde \Omega. \nonumber
\ea
In above equations $\tilde \xi^\alpha = (\xi^\alpha, \theta, \gamma)$, $G_\nu$ was implicitly defined and $\Omega$, accordingly with (\ref{vincintro}), is the constraint of the theory described by $\cl$ that is generated by $\nu$.

Proceeding as the symplectic formalism states, the constraint $\tilde \Omega$ (which will be called ''modified constraint") is added to the kinetic sector of $\tilde \cl_{\theta, \gamma}$ by a Lagrange multiplier, defining $\tilde \cl_{\theta, \gamma}^\1$, i. e.,
\be
	\tilde \cl_{\theta, \gamma}^\1 = a_\alpha \dot \xi^\alpha + (\Psi + \gamma) \dot \theta + \tilde \Omega \dot \lambda - \tilde V_{\theta, \gamma}.
\ee

One could modify $\tilde V_{\theta, \gamma}$ by using $\tilde \Omega = 0$, as suggested in \cite{b-w}, but we will not proceed in this way, so $\tilde V_{\theta, \gamma}$ still is $V + G + \frac 12 k \gamma \gamma$.

The symplectic matrix of $\tilde \cl_{\theta, \gamma}^\1$, being $\tilde \xi^{\1 \alpha} = ( \xi^\alpha, \theta, \gamma, \lambda)$, is 
\be
	\tilde f_{\theta, \gamma}^\1 (\vx,\vy) = \pmatrix{(f_{\alpha \beta}) & \frac{\delta \Psi(\vy)}{\delta \xi^\alpha(\vx)} & 0_\alpha & \frac{\delta (\Omega + G_\nu)(\vy)}{\delta \xi^\alpha(\vx)}\\[0.1in] \cr 
	- \frac{\delta \Psi(\vx)}{\delta \xi^\beta(\vy)} & \Theta_{xy} & -\dirac & \frac{\delta G_\nu(\vy)}{\delta \theta(\vx)}\\[0.1in] \cr 
	0_\beta & \dirac & 0 & kb(\vy) \dirac  \\[0.1in] \cr
	- \frac{\delta (\Omega + G_\nu)(\vx)}{\delta \xi^\beta(\vy)} & - \frac{\delta G_\nu(\vx)}{\delta \theta(\vy)} & - kb(\vx) \dirac & 0}.
\ee

	With this matrix we have more possibilities to the zero-modes. One of these is the following pair
\ba
	\label{intronugamma}
	\tilde \nu_\gamma^\alpha &=& \pmatrix{\nu^\alpha(\vx) & \;\;\; 0 & \;\; b & 0} = \pmatrix{\tilde \nu^\alpha(\vx) & 0}, \\
	\label{intronutheta}
	\tilde \nu_\theta^\alpha &=& \pmatrix{\mu^\alpha(\vx) &  -kb & 0 &  1}.
\ea
The $\mu^\alpha(\vx)$ is to be selected accordingly to the desired set of gauge generators.

	Before studding the conditions that come from imposing $\tilde \nu_\theta$ and $\tilde \nu_\gamma$ to be the zero-modes of $\tilde f^\1_{\theta,\gamma}$, it is easier to first demand that both supposed zero-modes do not generate new constraints. For $\tilde \nu_\gamma$ we find
\be
	0 = \int d^ny \; \tilde \nu_\gamma^\alpha(\vx) \frac {\delta \tilde V_{\theta,\gamma}(\vy)}{\delta \tilde \xi^{\1 \alpha}(\vx)} = \int d^ny \; \tilde \nu^\alpha(\vx) \frac {\delta \tilde V_{\theta,\gamma}(\vy)}{\delta \tilde \xi^\alpha(\vx)} \nonumber = \tilde \Omega(\vx).
\ee
Hence, no new condition has emerged for $G$, for we already knew that $\tilde \Omega = 0$.

	The function $G$ is obtained by demanding that $\tilde \nu_\theta$ does not produce new constraints, i.e.,
\ba
	0 & = & \int d^ny \; \tilde \nu_\theta^\alpha(\vx) \frac {\delta \tilde V_{\theta,\gamma}(\vy)}{\delta \tilde \xi^{\1 \alpha}(\vx)} \nonumber \\[0.1in]
	& = & \int d^ny \; \left ( \mu^\alpha(\vx) \frac{\delta (V + G)(\vy)}{\delta \xi^\alpha(\vx)} - kb \frac{\delta G(\vy)}{\delta \theta(\vx)} \right ).
\ea
From the last equation we can find $G$, which will enable us to compute $G_\nu$ (see (\ref{introGnu})), leaving $\Psi$ as the only unknown function in the symplectic matrix $\tilde f^\1_{\theta,\gamma}$.

	To finish the embedding procedure, $\Psi$ and the constant $k$ need to be found. This is accomplished though the requirements
\ba
	\int d^nx \; \tilde \nu_\gamma^\alpha (\vx) \tilde f^{\1 \theta, \gamma}_{\alpha \beta} (\vx, \vy) &=& 0, \\[0.1in]
	\int d^nx \; \tilde \nu_\theta^\alpha (\vx) \tilde f^{\1 \theta, \gamma}_{\alpha \beta} (\vx, \vy) &=& 0.
\ea

	The infinitesimal gauge generators are divided into two sets, one associated with $\theta$, that comes from $\tilde \nu_\theta$, and the other with $\gamma$, from $\tilde \nu_\gamma$. Summing these independent gauge transformations, one achieve the following one
\ba
	(\delta_{\vep_\gamma} + \delta_{\vep_\theta})\xi^\alpha (\vx) & = & \int d^ny \; [\vep_{\gamma} (\vy) \nu^\alpha (\vy) + \vep_{\theta} (\vy) \mu^\alpha (\vy)] \dirac, \nonumber \\
	(\delta_{\vep_\gamma} + \delta_{\vep_\theta})\theta (\vx) & = & - \vep_{\theta} (\vx) kb,  \nonumber \\
	(\delta_{\vep_\gamma} + \delta_{\vep_\theta})\gamma (\vx) & = & \vep_{\gamma} (\vx) b, \\
	(\delta_{\vep_\gamma} + \delta_{\vep_\theta})\lambda (\vx) & = & \vep_{\theta} (\vx). \nonumber
\ea 

	Under above transformations, $\tilde \cl^\1_{\theta,\gamma}$ is explicitly invariant, that is, no equation of movement is necessary to prove that $(\delta_{\vep_\gamma} + \delta_{\vep_\theta})\tilde \cl^\1_{\theta,\gamma} = 0$ \cite{mon}. If  we drop the term $\dot \lambda \tilde \Omega$ we get back to $\tilde \cl_{\theta,\gamma}$ and no physical property is lost, for one can always redo the symplectic algorithm and find $\tilde \Omega$ again as a constraint. So it is possible to use the equation of movement $\dot \theta = k \gamma$ that comes from $\tilde \cl_{\theta, \gamma}$. Using $\vep_\gamma = - \dot \vep_\theta \equiv \vep$,  $\; \delta_{\vep_\theta} + \delta_{\vep_\gamma} \equiv \delta_\vep$ and eliminating $\gamma$ from $\tilde \cl_{\theta, \gamma}$ we have
\ba
	\delta_\vep \xi^\alpha (\vx) & = & \int d^ny \; [- \dot \vep (\vy) \nu^\alpha (\vy) + \vep (\vy)\mu^\alpha (\vy)] \dirac,  \nonumber \\
	\label{introsymdotvep}
	\delta_\vep\theta (\vx) & = & - \vep (\vx) kb, \\
	\delta_\vep\dot \theta (\vx) & = & - \dot \vep (\vx) kb. \nonumber
\ea 

Note that in above set of generators appears the temporal derivative of the infinitesimal parameter, a feature which does not occured while working with $\tilde \cl_\theta$. This structure will be employed on the Proca model and the noncommutative self-dual model. As a special case, a St\"uckelberg-like Lagrangian will be achieved.

In next sections some applications and more details of this method of embedding will be shown.


\section{The Proca Model}
This first application will be specially useful to show how to deal with nonlinear-velocity  Lagrangians and how to achieve a St\"uckelberg-like Lagrangian, that is, a shift on $A^\mu$ that turns it into $A^\mu - \partial^\mu \theta$. To this end, we will look for gauge generators of the type $\;\;\; \delta_\vep A^\mu = \partial^\mu \vep \;\;\;$ and $ \;\;\; \delta_\vep \theta = \vep$. Actually, it will be done a more general embedding which has above structure as a special case. The presence of a temporal derivative acting on the infinitesimal parameter, as previously explained, will require the use of a Lagrangian of the type $\tilde \cl_{\theta, \gamma}$.

Before proceeding with the gauge embedding method, we will introduce the canonical momenta as new independent fields, turning the theory into a linear one (otherwise it would not be possible to use a symplectic framework). With the metric $g = \mbox{diag} \pmatrix {+ & - & - & -}$, the Proca Lagrangian,
\be
	{\cal L}(A_\mu,\partial_\nu A_\mu)  =  -\frac 14 F^{\mu \nu}F_{\mu \nu} + \frac {m^2}2 A^\mu A_\mu, 
	\label{proca1}
\ee
can be written as
\be
	{\cal L}(A_\mu,\partial_\nu A_i, \pi_i, \partial_j \pi_i) =  \pi^i \dot A_i + \frac 12 \pi^i \pi_i - \pi^i\partial_i A_0 - \frac 14 F^{ij}F_{ij} + \frac {m^2}2 A^\mu A_\mu,
	\label{proca2}
\ee
where $\mu=0,1,2,3$, $i=1,2,3$ and $F_{\mu \nu} \equiv \partial_\mu A_\nu - \partial_\nu A_\mu$. Through the Euler-Lagrange equations of last Lagrangian, one can find that $\pi_i = F_{i 0}$. If $\pi_i$ is replaced by $F_{i0}$ in last Lagrangian, one gets back to the first one.

The next step is to introduce the Wess-Zumino fields $\theta$ and $\gamma$ (just like (\ref{ltg})):
\be
	\tilde {\cal L} = \pi_i \dot A^i + (\Psi + \gamma) \dot \theta - \tilde V,
\ee 
with
\be
	\tilde V \equiv  - \frac 12 \pi^i \pi_i + \pi^i \partial_i A_0 + \frac 14 F^{ij}F_{ij} - \frac {m^2}2 A^\mu A_\mu + G  + \frac k2 \gamma \gamma .
	\label{vproca}
\ee
The constant k as well as the functions $G$ and $\Psi$ are still unknown, but, by definition, $G$ is zero when $\theta$ is zero (the unitary gauge) and both functions depend on $A_\mu, \pi_i, \theta$ and its spatial derivatives, which ensures that $\dot \theta = k \gamma$. In order to find the gauge embedded Lagrangian, all our work resides on fixing $k$ and finding the functions $\Psi$ and $G$.  

Let $\tilde \xi^\alpha = (A^0, A^i, \pi^i, \theta, \gamma)$ be the symplectic coordinates, then  $\tilde a_\alpha = (0, \pi_j, 0, \Psi + \gamma, 0)$ are the symplectic momenta (see (\ref{LL})) and 
\be
	\tilde f = \pmatrix{ 0 & 0 & 0 & \frac{\delta \Psi(\vy)}{\delta A^0(\vx)} & 0 \cr
	0 & 0 & -g_{ji} \dirac & \frac{\delta \Psi(\vy)}{\delta A^i(\vx)} & 0 \cr
	0 & g_{ij} \dirac & 0 & \frac{\delta \Psi(\vy)}{\delta \pi^i(\vx)} & 0 \cr
	- \frac{\delta \Psi(\vx)}{\delta A^0(\vy)} & - \frac{\delta \Psi(\vx)}{\delta A^j(\vy)} & - \frac{\delta \Psi(\vx)}{\delta \pi^j(\vy)} & \Theta_{xy} & - \dirac \cr
	0 & 0 & 0 & \dirac & 0}
\ee
is the symplectic matrix, whose components are\cite{f-j}
\be
	\tilde f_{\alpha \beta} (\vx,\vy) \equiv \frac{\delta \tilde a_\beta(\vy)}{\delta \tilde \xi^{\alpha}(\vx)} - \frac{\delta \tilde a_\alpha(\vx)}{\delta \tilde \xi^{\beta}(\vy)}.
\ee
The symbol $\Theta_{xy}$ is just a shorthand notation for $\delta \Psi(\vy)/\delta \theta (\vx) - \delta \Psi(\vx)/\delta \theta (\vy)$. Note that $\tilde f$ is a $9 \times 9$ matrix with two space indexes in each entry. There is also an implicit time dependence, which comes from the coordinates and momenta. In above representation of $\tilde f$, some zeros in it are actually null columns, null lines or null matrixes. 

In the symplectic framework, a theory has gauge symmetry if, and only if, the symplectic matrix is degenerate and its zero-modes does not produce new constraints\cite{b-w, mon}. In that case, the components of the zero-modes will be the infinitesimal gauge generators. Although $\Psi$ is still an arbitrary function, the presence of the Dirac deltas in last column and last line severely restraint our possibilities of choosing zero-modes (as explained in last section). With the purpose of avoiding such restraint on our choices, before trying to gauge embed the theory, we will insert a constraint in the Lagrangian.

In order to generate a suitable new constraint, let us demand
\be
	\tilde \nu  = \pmatrix {1 & 0_{1 \times 3} & 0_{1 \times 3} & 0 & b} 
\ee
to be the zero-mode of $\tilde f$, where $b$ is a constant. The Proca model (\ref{proca2}), whose symplectic matrix is the one above without last two lines and columns, has the zero-mode $\nu = \pmatrix{1 & 0_{1\times 3} & 0_{1 \times 3}}$, hence $\tilde \nu$ complies with $\tilde \nu = \pmatrix{ \nu & 0 & b}$, being in accordance with (\ref{introzmmod}). 

	The constraint generated by $\nu$ in the Proca model is $\Omega = -\partial_i\pi^i - m^2 A_0$. As we will see, $\tilde \nu$ will produce a constraint which is equal to $\Omega$ when $\gamma = \theta = 0$.

	Demanding $\tilde \nu$ to be a zero-mode of $\tilde f$, one condition for $\Psi$ is found, which is
\be
	\frac {\delta \Psi(\vy)}{\delta A^0(\vx)} = -b \dirac.
	\label{psi_a0proca}
\ee

As well known from the symplectic theory, the constraint emerge from the following contraction:
\be
	\tilde \Omega (\vx) = \int d^3y \; \tilde \nu^\alpha(\vx) \frac{\delta \tilde V(\vy)}{\delta \tilde \xi^\alpha (\vx)} = -\partial_i \pi^i - m^2A_0 + \int d^3y \; \frac{\delta G(\vy)}{\delta A_0 (\vx)}  + bk \gamma,
	\label{modconstproca}
\ee
or, for short, $\tilde \Omega = \Omega + G_0 + bk\gamma$, being $G_0$ implicitly defined.

Following the standard procedure for handling constraints in the symplectic framework \cite{b-w}, we add $\dot \lambda \tilde \Omega$ to $\tilde \cl$ and treat $\lambda$ as a new independent field, that is, a Lagrange multiplier. Hence,
\be
	\tilde \cl^{(1)} = \pi^i \dot A_i + (\Psi + \gamma) \dot \theta + \dot \lambda \tilde \Omega - \tilde V.
\ee
The presence of the constraint in the kinetic part of the Lagrangian allow us to remove it from the potential part. Nevertheless, this common procedure would be of no help here, therefore no change was done in the potential.

Setting $\tilde \xi^{(1)\alpha} = (A^0, A^i, \pi^i, \theta, \gamma, \lambda)$ as the new symplectic coordinates, where hereafter $\alpha = 1,2,...,10$, and with the help of equation (\ref{psi_a0proca}), the following symplectic matrix is achieved:

\be
\tilde f^{(1)} = \pmatrix{ 0 & 0 & 0 & -b\delta^{(3)} & 0 & \frac {\delta G_0(\vy)}{\delta A_0(\vx)} - m^2\delta^{(3)} \\[0.1in] \cr
	0 & 0 & -g_{ji} \delta^{(3)} & \frac{\delta \Psi(\vy)}{\delta A^i(\vx)} & 0 & \frac {\delta G_0(\vy)}{\delta A^i(\vx)} \\[0.1in] \cr
	0 & g_{ij} \delta^{(3)} & 0 & \frac{\delta \Psi(\vy)}{\delta \pi^i(\vx)} & 0 & \frac {\delta G_0(\vy)}{\delta \pi^i(\vx)} - {\partial}_i^y \delta^{(3)} \\[0.1in] \cr
	b\delta^{(3)} & - \frac{\delta \Psi(\vx)}{\delta A^j(\vy)} & - \frac{\delta \Psi(\vx)}{\delta \pi^j(\vy)} & \Theta_{xy} & - \delta^{(3)} & \frac {\delta G_0(\vy)}{\delta \theta (\vx)} \\[0.1in] \cr
	0 & 0 & 0 & \delta^{(3)} & 0 & bk \delta^{(3)} \\[0.1in] \cr
	- \frac {\delta G_0(\vx)}{\delta A_0(\vy)} + m^2 \delta^{(3)} & \frac {\delta G_0(\vx)}{\delta A^j(\vy)} & {\partial}_j^x \delta^{(3)} - \frac {\delta G_0(\vx)}{\delta \pi^j(\vy)} & - \frac {\delta G_0(\vx)}{\delta \theta(\vy)} & -bk \delta^{(3)} & 0}.
\ee
For the sake of clarity, it was convenient to use the notation $\delta^{(3)}$ instead of $\dirac$.

Now we are in position to choose the symmetry the embedded theory will have. In accordance with (\ref{intronugamma}) and (\ref{intronutheta}),  we can select to independent zero-modes to become the infinitesimal gauge generators, which are
\ba
	\tilde \nu_{(\theta)} & = & \pmatrix{a_0 & a \partial^i & c \partial^i & -kb & 0 & 1}, \nonumber \\
	\tilde \nu_{(\gamma)} & = & \pmatrix{1 & 0_{1 \times 3} & 0_{1 \times 3} & 0\;\; & b\; & 0} = \pmatrix{ \tilde \nu & 0}.
	\label{zmproca}
\ea
The values of the constants $a_0, \; a$ and $c$ can be freely selected, remembering that different choices directly correspond to different gauge generators (see \ref{introsymdotvep}). As it will be shown, the value of $b$ is also free. 

Naturally, other structures of zero-modes are possible, some of which entail correspondence to both Wess-Zumino fields in each set; and, instead of just constants or spatial derivatives, dependence on $A^\mu$ or $\pi^i$, for example, is also possible. The consequences of such alternatives approaches are still a subject to be studied.

Although at this stage we could fix above mentioned constants, selecting some of them to be zero,   simplifying considerably our work, we will deal with the problem in the present general form, disclosing a wider symmetry.

For $\tilde \nu_{(\gamma)}$ just one condition is necessary to assure its zero-mode nature, namely,
\be
	\frac{\delta G_0 (\vy)}{\delta A_0 (\vx)} = (m^2-b^2k)\dirac.
	\label{G0proca}
\ee

That zero-mode need to be a generator of gauge symmetry, therefore no new constraint may arise from its contraction with the gradient of the potential. By equations (\ref{modconstproca}) and (\ref{zmproca}), we see that this wish was automatically fulfilled.

Now lets turn our attention to $\tilde \nu_{(\theta)}$. There is a set of nontrivial equations that need to be satisfied in order to $\tilde \nu_{(\theta)}$ be a zero-mode of $\tilde f^{(1)}$. Instead of evaluating them now, it seems to be easier to first demand that $\tilde \nu_{(\theta)}$ may not give rise to a new constraint. Hence,
\ba
	0 & = & \int d^3y \; \tilde \nu^\alpha_{(\theta)} (\vx) \; \frac{\delta \tilde V (\vy)}{\delta \tilde \xi^{(1) \alpha}(\vx)} \nonumber \\
	\label{Gproca}
	& = & \int d^3y \left \{ a_0 \dirac ( -\partial_i \pi^i - m^2A_0 )  + a\partial^i_x \dirac ( \partial^j F_{ij} - m^2 A_i )  +     \right. \\
	& & \left. + \; c\partial^i_x  \dirac ( -\pi_i + \partial_i A_0 ) + \rho^\mu_x \frac{\delta G(\vy)}{\delta A^\mu(\vx)} + c\partial^i_x \frac{\delta G(\vy)}{\delta \pi^i(\vx)} - kb \frac{\delta G(\vy)}{\delta \theta(\vx)} \right \}. \nonumber
\ea
The index $x$ on $\partial^i$ means that the derivative must be evaluated with respect to $x$ (i.e., $\partial_x^i \equiv \partial / \partial x_i$), and 
\be
	\rho^\mu_x \equiv ( a_0, a\partial^i_x)
\ee

	Equation (\ref{Gproca}) can be solved by treating $G$ as a power series of $\theta$ (and its spatial derivatives). Let $\cg_n$ be proportional to $\theta^n$, so $G = \sum_n \cg_n$. The condition $G(\theta = 0) = 0$ leads to $n \ge 1$. Hence,
\be
	\cg_1 = \frac{\theta}{kb} ( -a_0 \partial_i\pi^i -m^2\rho^\mu A_\mu - c\partial^i\pi_i + c\partial^i\partial_i A_0).
\ee

The terms $\frac{\delta G(\vy)}{\delta A^\mu(\vx)}$ and $\frac{\delta G(\vy)}{\delta \pi^i(\vx)}$ do not contribute to the computation of $\cg_1$, but them do contribute to others $\cg_n$'s, for they are the ones, besides $\frac{\delta G(\vy)}{\delta \theta(\vx)}$, that enclose the $\theta$ field. After some straightforward calculations, one can find $\cg_2$ (without surface terms) as
\be
	\cg_2 = - \frac 1{2(kb)^2} \{ c(2 a_0 + c) \partial_i \theta \partial^i \theta  + m^2 \rho^\mu \theta \rho_\mu \theta \}.
\ee 

The absence of $A^\mu$ and $\pi^i$ in $\cg_2$ implies $\cg_n = 0$ for all $n \ge 3$. Thus the function $G$ is completely known, and we can write down the expression for $G_0$, that is,
\be
	G_0(\vx) \equiv \int d^3y \; \frac {\delta G(\vy)}{\delta A^0 (\vx)} = \frac 1 {kb} (c\partial^i \partial_i \theta - m^2 a_0 \theta).
	\label{G0proca2}
\ee
Applying this result in equation (\ref{G0proca}), we have
\be
	k = \frac {m^2}{b^2}.
	\label{kproca}
\ee
This fixes $k$ in relation to $b$.

	Our next and final step in order to prescribe the gauge embedded Lagrangian is to find $\Psi$. This can be done by demanding $\tilde \nu_{(\theta)}$ to be a zero-mode of $\tilde f^{(1)}$. Among some redundant or trivial equations, emerge the following important ones (using (\ref{G0proca2}) and (\ref{kproca})): 
\be
	c\partial_j^x \dirac + \frac{m^2}{b} \frac{\delta \Psi(\vx)}{\delta A^j(\vy)} = 0,
	\label{psiaproca}
\ee
\be
	- a\partial_j^x \dirac + \frac {m^2}b \frac {\delta \Psi (\vx)}{\delta \pi^j(\vy)} + \partial_j^x \dirac = 0,
	\label{psipiproca}
\ee
\be
	-b a_0 \dirac + a\partial_x^i \frac{\delta \Psi(\vy)}{\delta A^i(\vx)} + c \partial^i_x \frac {\delta \Psi (\vy)}{\delta \pi^i(\vx)} - \frac {m^2}b \Theta_{xy} - \frac{\delta G_0 (\vx)}{\delta \theta (\vy)} = 0.
	\label{psithetaproca}
\ee
	With equations (\ref{psi_a0proca}) and (\ref{psiaproca}-\ref{psithetaproca}), up to an additive function just of $\theta$ (action surface term), $\Psi$ can be determined. The answer is
\be
	\Psi = - \frac b {m^2} \{ m^2 A_0 + c \partial_i A^i + (1 - a)\partial^i \pi_i \}.
\ee

Gathering all was done, the gauge version of the Lagrangian of the Proca model was found. Nevertheless, it is more interesting to express it without the momenta $\pi_i$. At first, note that we can drop the term $\dot \lambda \tilde \Omega$ from $\tilde \cl^{(1)}$ without changing the dynamic (one can always redo the symplectic algorithm and find again the constraint $\tilde \Omega$), this will lead us back to $\tilde \cl$. By varying $\tilde \cl$ with respect to $\pi_i$ and using Euler-Lagrange equations we find
\be
	\pi_i = \partial_i A_0 - \dot A_i + \frac b {m^2} \{(1-a) \partial_i \dot \theta + (a_0 + c) \partial_i \theta \}.
\ee

Note that the momenta are not the original ones (which are $F_{i0}$), but when $\theta$ is removed they are recovered.

Also from the Euler-Lagrange equations, we have
\be
	\gamma = \frac {b^2}{m^2} \dot \theta.
\ee

Thus, eliminating $\pi_i$ and $\gamma$, the Lagrangian $\tilde \cl$ can be expressed by
\ba
	\tilde \cl & = & - \frac 14 F_{\mu \nu} F^{\mu \nu} + \frac {m^2}2 A^\mu A_\mu + \frac b {m^2} \{ -m^2 A_0 \dot \theta + (1-a) \partial_i \dot \theta (\partial^i A_0 - \dot A^i) + a_0 \theta (\partial^i \partial_i A_0 - \partial_i \dot A^i ) + \theta m^2 \rho^\mu A_\mu \} + \nonumber \\[0.1in]
	\label{procaembedL}
	&& + \frac {b^2}{m^4} \left \{ \frac 32 (1-a)^2 \partial_i \dot \theta \partial^i \dot \theta - \frac 12 a_0^2 \partial_i \theta \partial^i \theta + (1-a) (a_0 + c) \partial_i \dot \theta \partial^i \theta + \frac{m^2}2 \rho^\mu \theta \rho_\mu \theta \right \} + \frac {b^2}{2m^2}\dot \theta \dot \theta.
\ea

	From the components of $\tilde \nu_\theta$ and $\tilde \nu_\gamma$ the infinitesimal gauge generators of the theory are obtained as (see (\ref{introsymdotvep}))
\ba
	\delta_\vep A_0 & = & \vep a_0 -  \dot \vep, \nonumber \\
	\delta_\vep A^i & = & - a \partial^i \vep,  \\
	\delta_\vep \theta & = & - \frac {m^2}b \vep. \nonumber
\ea

	The symplectic formalism assures us that $\tilde \cl$ is invariant under the above transformations for any constants $b$, $a_0$ and $a$ (assuming they have proper dimensions, which are squared mass, mass and unit respectively). 

	Usually, terms with more than two derivatives in the Lagrangian are not wanted, these can be avoided by fixing $a = 1$.

	If one wants an explicit Lorentz invariance, the constants need be fixed as $b= m^2$, $a = 1$ and $a_0 = 0$ (alternatively, $b$ could also be $-m^2$). With these values, the Lagrangian turn out to have a St\"uckelberg aspect, that is
\be
	\tilde \cl  =  - \frac 14 F_{\mu \nu} F^{\mu \nu} + \frac {m^2}2 A^\mu A_\mu - m^2A^\mu \partial_\mu \theta + \frac {m^2}2 \partial^\mu \theta \partial_\mu \theta.
\ee

This result could also be achieved by analyzing the gauge generators and comparing them to (\ref{introsymdotvep}), knowing that the wanted generators are $\delta_\vep A^\mu = - \partial^\mu \vep \;$ and $\; \delta_\vep \theta = -\vep \;\;$ (in order to $\delta_\vep(A^\mu + \partial^\mu \theta)=0$). Doing so, one could fix the constants as soon as they have appeared, achieving above results more quickly.

	The Lagrangian of equation (\ref{procaembedL}) if not the most general one that can be achieved with the symplectic embedding method. Others structures of the zero-modes $\tilde \nu_{(\theta)}$ and $\tilde \nu_{(\gamma)}$ are also possible, and their components, together with the components of $\tilde \nu$, could also be field dependent.


\section{Irrotational Fluid Model}

	In this section, the symplectic embedding formalism will be applied on an unconstrained theory. To this end it is necessary to use a Lagrangian of the type $\tilde \cl_\theta$ (\ref{introLtheta}).

	The irrotational fluid model has its dynamics governed by the following Lagrangian density
\begin{equation}
\label{00010}
{\cal L} = - \rho\dot\eta + \frac 12 \rho(\partial_a\eta)(\partial^a\eta) - \frac g\rho,
\end{equation}
where $a=1,2,...,d$ (runs through spatial indexes only), $\;\; \rho$ is the mass density, $\eta$ is the velocity potential and $g$ is a constant. Here the metric is Euclidean. This model does not possess neither gauge symmetry nor, contrary to the previous model, constraints in the symplectic sense \footnote{The Dirac algorithm always set constraints to any linear Lagrangian, but, for each constraint, it adds a new field: the canonical momenta. This roundabout procedure is not present in the symplectic algorithm. See \cite{f-j,mon} for details.}. The above Lagrangian is already linear in the velocity, hence we can proceed directly to the embedding process. 

In accordance with last comments, we will not use the $\gamma$ field. Doing so, the gauge embedded Lagrangian has the aspect
\be
\label{00020}
{\tilde{\cal L}} = - \rho\dot\eta + \Psi \dot \theta + \frac 12 \rho \; \partial_a\eta\partial^a\eta - \frac g\rho - G,
\ee
where $\Psi\equiv \Psi(\rho, \eta, \theta)$ and $G \equiv G(\rho, \eta, \theta)$. 

Setting the symplectic coordinate vector as $\tilde \xi^{ \alpha} = (\rho, \eta, \theta)$, the symplectic momenta and matrix are \linebreak $\tilde a_\alpha = (0, -\rho, \Psi)$ and
\be
\label{00030}
{\tilde f }= \pmatrix{0 & - \delta(\vec x - \vec y) & \frac{\delta\Psi(\vec y)}{\delta\rho(\vec x)} \\[0.1in] \cr
\delta(\vec x - \vec y) & 0 & \frac{\delta\Psi(\vec y)}{\delta\eta(\vec x)}  \\[0.1in] \cr
- \frac{\delta\Psi(\vec x)}{\delta\rho(\vec y)} & -  \frac{\delta\Psi(\vec x)}{\delta\eta(\vec y)} & \Theta_{xy}  }.
\ee
It is worth to remark that $\vx$ and $\vy$ are $d$-dimensional vectors and, like previous application, $\Theta_{xy} \equiv \delta \Psi (\vy)/ \delta \theta (\vx) - \delta \Psi (\vx)/ \delta \theta (\vy)$.

The symplectic method states that if the symplectic matrix is degenerated and one of its (linearly independent) zero-modes does not produce any new constraints, then the theory has gauge symmetry and the infinitesimal gauge generators are given by the components of that zero-mode. Due to the absence of the $\gamma$ field, there is no ''modified constraint" to insert, so we can go right to the selection of the zero-mode related to the infinitesimal gauge generators. The most general constant zero-mode of $\tilde f^{(0)}$ has the form
\be
	\tilde \nu = \pmatrix {a & b & 1}.
\ee
This one imposes the following conditions on $\Psi$:
\ba
	\frac {\delta \Psi (\vx)}{\delta \rho(\vy)} & = &  b \dirac \nonumber \\
	\label{fluidpsieqs}
	\frac {\delta \Psi (\vx)}{\delta \eta(\vy)} & = & - a \dirac\\
	\Theta_{xy} & = & 0 \nonumber.
\ea

	If one is not interested on a gauge symmetry related to $\rho$ (i.e., $\delta_\vep \rho = 0$), for example, $a$ could be set equal to zero at this point, simplifying the calculations to come.

{}From equations (\ref{fluidpsieqs}) it is possible to find $\Psi$ as
\be
	\Psi = b \rho - a \eta + f(\theta),
\ee
where $f(\theta)$ is an arbitrary function of $\theta$ alone. This function, as one can easily check, only contributes to a surface term to the action, therefore it will not be written anymore.

The last step to gauge embed this theory is the calculation of $G$. This function can be found by demanding that $\tilde \nu$ does not gives rise to any constraint, that is,
\be
\label{condGfluid}
\int d^d y\,\,\tilde \nu^\alpha(\vec x) \frac{\delta \tilde V(\vec y)}{\delta \tilde \xi^\alpha(\vec x)} = 0,
\ee
with $\tilde V$ being the potential part of $\tilde {\cal L}$, namely,
\be
	\tilde V = - \frac 12 \rho \; \partial_a\eta\partial^a\eta + \frac g\rho + G.
\ee
Hence
\be
\label{00090}
\int d^d y\,\,\left\{ a \left ( - \frac 12\partial_a\eta \; \partial^a\eta \; \dirac - \frac g {\rho^2} \dirac + \frac {\delta G (\vy)}{\delta \rho (\vx)} \right ) + b \left ( - \rho \; \partial_a \eta \; \partial^a \dirac + \frac {\delta G (\vy)}{\delta \eta (\vx)} \right ) + \frac {\delta G (\vy)}{\delta \theta (\vx)} \right\}=0.
\ee
In above equation, every implicit dependence on space refers to the vector $\vy$.

Expanding $G$ in powers of $\theta$, $G = \sum \cg_n$ with $\cg_n \propto \theta^n$ and $n \ge 1$ (due to $G(\theta=0)=0$), we have
\ba
\label{00100}
	\cg_1 &=& a \left ( \frac 12 \partial_a\eta \; \partial^a\eta \; \theta + \frac g {\rho^2} \theta \right ) + b \rho \partial_a \eta \; \partial^a \theta, \nonumber\\[0.1in]
	\cg_2 &=&  - a \left ( -a \frac g {\rho^3} \theta^2 + b \partial^a \eta \; \partial_a \theta \; \theta \right ) - \frac {b^2 }2 \rho \partial_a \theta \; \partial^a \theta, \nonumber\\[0.1in]
	\cg_3 & = & a \left ( a^2 \frac g {\rho^4} \theta^3 + \frac {b^2} 2\theta \partial^a \theta \; \partial_a \theta \right ), \\[0.1in]
	\cg_n & = & a^n \frac {g}{\rho^{n+1}} \theta^n \;\;\;\;\;\;\; \forall \;\; n \ge 4. \nonumber 
\ea

Being $\rho > a \theta$ the series $\sum \cg_n$ converges, and we find the following Lagrangian: 
\be
	\tilde \cl = - \rho \dot \eta + (b \rho - a \eta) \dot \theta +  (\rho - a\theta) \left ( \frac 12 \partial_a \eta \; \partial^a \eta - b \partial^a \eta \; \partial_a \theta + \frac {b^2}2  \partial^a \theta \; \partial_a \theta \right ) - \frac g {\rho - a \theta}.
\ee 

The above Lagrangian is invariant under gauge transformations that result from (\ref{introLthetagen})
\ba
\label{00150}
\delta_\vep \rho &=& a \vep,\nonumber\\
\delta_\vep \eta &=& b \vep,\\
\delta_\vep \theta&=& \vep. \nonumber
\ea

	One can easily check that $\delta_\vep \tilde \cl = 0$ (for $\delta_\vep$ acts like a derivative operator).


\section{Noncommutative Self-Dual Model}

This section contains our main result, the extension of the symplectic embedding into the noncommutative scenario.
By means of some results from the symplectic formalism we construct a dual theory, with gauge symmetry, to
the noncommutative self-dual model in 2+1 dimensions. We use the
facilities of the symplectic formalism for handling infinitesimal
gauge generators in order to obtain some generality on our final
Lagrangian, which has, as a special case, a St\"uckelberg aspect.
The duality is established without the use of the Seiberg-Witten
map and with no restriction on the powers of the parameter of the
Moyal Product.

Systematically we attain a Lagrangian with gauge symmetry and the
same ''physics" of the noncommutative self-dual model, without
resorting to any kind of approximation or restriction to the Moyal
product. Deliberately some of its parameters are left unfixed;
because, doing so, one can analyze the gauge generators of this
Lagrangian, compare these with the desired ones and fix the
parameters accordingly. At the end of this letter, these
parameters are fixed with the aim of achieving a Lagrangian with a
St\"uckelberg aspect.

\subsection{Setting the problem}

To achieve our objective, what we need is a modified noncommutative self-dual Lagrangian whose symplectic matrix is degenerated and its zero-modes do not produce new constraints. Nevertheless, this new Lagrangian, with some gauge fixing conditions (the unitary gauge), must be equal to the original one (except for surface terms).

The noncommutative self-dual Lagrangian is
\be
	{\cal L} = \frac 12 f^{\mu} f_{\mu} - \frac 1{4m} \epsilon^{\mu \nu \lambda} f_\mu F_{\nu \lambda},
\ee
with summation convention implied, $\mu, \nu, \lambda = 0,1,2$, metric $g= \mbox{diag} \pmatrix{+ & - & -}$ and $\epsilon^{012} = 1$. The noncommutative field nature of this Lagrangian resides solely in 
\be
	F_{\mu \nu} = \partial_\mu f_\nu - \partial_\nu f_\mu - ie[f_\mu,f_\nu],
\ee
where $[,]$ is the Moyal commutator, that is,
\be
	[f_\mu, f_\nu](\vx) = (f_\mu \star f_\nu)(\vx) - (f_\nu \star f_\mu)(\vx),
\ee
and the Moyal product is defined by \cite{nc}
\be
(f_\mu \star f_\nu)(\vx) \equiv e^{\frac i2 \theta_{ij}  \partial^i_x  \partial^j_y}f_\mu(\vx) f_\nu(\vy)|_{\vy \rightarrow \vx}.
\ee

In order to use some symplectic results, kinetic and potential parts of $\cal L$ need to be separated. This Lagrangian can be written as
\be
	{\cal L} = \frac 1{2m} \epsilon^{ij} f_i \dot f_j - V,
\ee
where $i,j = 1,2$, $\epsilon^{12}=1$ and
\be
	V = -\frac 12 f^\mu f_\mu + \frac 1m \ep^{ij}f_i\partial_jf_0 - \frac 3{4m} f_0 \ep^{ij}ie[f_i,f_j].
\ee

Now we will introduce two Wess-Zumino fields ($\theta$ and $\gamma$) and two unknown functions, defining $\tilde \cl$:
\be
	\tilde {\cal L} (f_\mu, \dot f_\mu, \theta, \dot \theta, \gamma) = \frac 1{2m} \epsilon^{ij} f_i \dot f_j + (\Psi + \gamma)\dot \theta - \tilde V,
\ee
where
\ba
	\Psi &\equiv& \Psi(f_\mu, \theta), \nonumber \\
	\tilde V & \equiv & V + G + \frac 12 k \gamma \gamma, \\
	G &\equiv& G(f_\mu, \theta) \nonumber
\ea
and $k$ is a constant. The dependence on the spatial derivatives is implicit in above equations. The function $G$ satisfies the condition $G(\theta =0)=0$.

\subsection{The modified constraint}
The Lagrangian $\tilde \cl$ is not supposed to be explicitly invariant under some set of gauge transformations; as it will be shown, a constraint must be added to this end.

Let
\ba
	(\tilde \xi^{ \alpha} ) & = & \pmatrix{f_0 & f^i & \theta & \gamma}, \nonumber \\
	(\tilde a_\alpha) & = & \pmatrix{0 & \frac 1{2m} \ep_{ij}f^i & \Psi + \gamma & 0}
\ea
be the symplectic coordinates and momenta respectively, with $\alpha = 1,2,...,5$. Thus, the symplectic matrix,
\be
	\tilde f_{\alpha \beta} (\vx,\vy) \equiv \frac{\delta \tilde a_\beta(\vy)}{\delta \tilde \xi^{\alpha}(\vx)} - \frac{\delta \tilde a_\alpha(\vx)}{\delta \tilde \xi^{\beta}(\vy)},
\ee
is given by
\be
\tilde f_{\alpha \beta} = \pmatrix{ 0 & 0 & \frac{\delta \Psi(\vy)}{\delta f_0(\vx)} & 0 \cr
0 & \frac {\ep_{ij}}m \delta (\vx - \vy) & \frac{\delta \Psi(\vy)}{\delta f^i(\vx)} & 0 \cr
- \frac{\delta \Psi(\vx)}{\delta f_0(\vy)} & -\frac{\delta \Psi(\vx)}{\delta f^j(\vy)} & \Theta_{xy} & -\delta (\vx -\vy) \cr
0 & 0 & \delta (\vx - \vy) & 0},
\ee
where, as before,
\be
\Theta_{xy} \equiv \frac{\delta \Psi(\vy)}{\delta \theta(\vx)} - \frac{\delta \Psi(\vx)}{\delta \theta(\vy)}.
\ee

Accordingly with (\ref{introzmmod}), let 
\be
	(\tilde \nu^{\alpha}) = \pmatrix{ 1 & 0_{1 \times 2} & 0 & b},
	\label{nut0}
\ee
with $b$ constant, be the zero-mode. Except for the last two components, $(\tilde \nu^{\alpha})$ is the zero-mode of the symplectic matrix of $\cl$. The choice done in last equation implies the following condition on $\Psi$:
\be
	\label{psi0}
	\frac{\delta \Psi (\vy)}{\delta f_0(\vx)} = -b \delta(\vx - \vy).
\ee
And, with that zero-mode, we find the constraint
\ba
\tilde \Omega(\vx) & \equiv & \int d^2y \; \tilde \nu^{\alpha} \frac {\delta \tilde V(\vy)}{\tilde \xi^{\alpha}(\vx)} \nonumber \\
	& = & -f_0(\vx) + \frac 1m \ep^{ij} \partial_i f_j(\vx) - \frac 3{4m} \ep^{ij}ie[f_i,f_j](\vx) + \int d^2y \frac{\delta G(\vy)}{\delta f_0(\vx)} + bk\gamma (\vx).
\ea
Using $\Omega$ to express the constraint of the original theory and $G_0(\vx) \equiv \int d^2y \frac{\delta G(\vy)}{\delta f_0(\vx)}$, we can write
\be
	\tilde \Omega = \Omega + G_0 + bk \gamma.
	\label{omegatil}
\ee

Following the symplectic approach, let us insert this constraint into the kinetic part of the Lagrangian $\tilde \cl$. Hence, we get
\be
	\label{l1}
	\tilde \cl^\1 = \frac 1{2m} \ep^{ij} f_i \dot f_j + (\Psi + \gamma) \dot \theta + \tilde \Omega \dot \lambda - \tilde V.
\ee

\subsection{The Generators of Gauge Transformations}
With $\alpha = 1,2,...,6$ and
\ba
	(\tilde \xi^{\1 \alpha} ) & = & \pmatrix{f_0 & f^i & \theta & \gamma & \lambda}, \nonumber \\
	(\tilde a_\alpha^\1) & = & \pmatrix{0 & \frac 1{2m} \ep_{ij}f^i & \Psi + \gamma & 0 & \tilde \Omega},
\ea
the symplectic matrix is
\be
\left(\tilde f^\1_{\alpha \beta}\right) = \pmatrix{ 0 & 0 & \frac{\delta \Psi(\vy)}{\delta f_0(\vx)} & 0 & \frac {\delta \tilde \Omega(\vy)}{\delta f_0(\vx)} \\[0.1in] \cr
0 & \frac {\ep_{ij}}m \delta (\vx - \vy) & \frac{\delta \Psi(\vy)}{\delta f^i(\vx)} & 0 & \frac {\delta \tilde \Omega(\vy)}{\delta f^i(\vx)} \\[0.1in] \cr
- \frac{\delta \Psi(\vx)}{\delta f_0(\vy)} & -\frac{\delta \Psi(\vx)}{\delta f^j(\vy)} & \Theta_{xy} & -\delta (\vx -\vy) & \frac {\delta G_0(\vy)}{\delta \theta (\vx)} \\[0.1in] \cr
0 & 0 & \delta (\vx - \vy) & 0 & kb\delta(\vx - \vy) \\[0.1in] \cr
- \frac {\delta \tilde \Omega(\vx)}{\delta f_0(\vy)} & -\frac {\delta \tilde \Omega(\vx)}{\delta f^j(\vy)} & -\frac {\delta G_0(\vx)}{\delta \theta (\vy)} & -kb \delta (\vx - \vy) & 0},
\ee

The zero-modes of $(\tilde f^\1_{\alpha \beta})$, which will be chosen, will turn out to be the gauge generators of the embedded theory. The structure of zero-modes selected in this work is
\ba
	(\tilde \nu_\theta^{\alpha}(\vx)) & = & \pmatrix{\rho_0 & \rho^i_x  & -kb & 0 & 1}, \nonumber \\
	(\tilde \nu_\gamma^{ \alpha}) & = & \pmatrix{1 & 0_{1 \times 2} & 0 & b & 0}.
\ea
As in the Proca model, our notation is such that the component $\rho_0$ is a constant and $\rho_x^i \equiv a \partial_x^i$, where $a$ is another constant. So, at this point, we have not yet fixed totally the gauge generators: $k, b, \rho_0$ and $\rho_x^i$ still present some freedom. A relation between $k$ and $b$ will be found but there is no other restriction; the final answer will be quite general, and, as a special case, we will find a St\"{u}ckelberg-like embedded theory. 

\subsection{The function $G$}
Lets assume the existence of a function $\Psi$ compatible with the zero-modes $\tilde \nu_\theta$ and $\tilde \nu_\gamma$. These need to be gauge generators, so the function $G$ must agree with
\be
	\int d^2y \tilde \nu_\theta^{\alpha} \frac{\delta \tilde V(\vy)}{\delta \tilde \xi^{\1 \alpha}(\vx)} = 	\int d^2y \tilde \nu_\gamma^{ \alpha} \frac{\delta \tilde V(\vy)}{\delta \tilde \xi^{\1 \alpha}(\vx)} = 0.
	\label{contrac}
\ee

For $\tilde \nu_\gamma$ there is no difficulty, its contraction with the gradient of $\tilde V$ is equal to $\tilde \Omega$, which is zero, accordingly to the kinetic part of $\tilde \cl^\1$. For $\tilde \nu_\theta$ we get the following differential equation:
\ba
	\int d^2y \; &&\left \{ \rho_0 \( \Omega(\vy) \delta(\vx -\vy) + \frac {\delta G(\vy)}{\delta f_0(\vx)} \) + \rho_x^i \dirac \( -f_i(\vy) + \frac 1m \ep_{ij} {\partial^j_y} f_0 (\vy) - \right. \right. \nonumber\\ 
&& \left. \left. - \frac {3ie}{2m} \ep_{ij}[f^j,f_0](\vy) \) + \rho_x^i \frac{\delta G(\vy)}{\delta f^i(\vx)} - kb \frac{\delta G(\vy)}{\delta \theta(\vx)} \right \} = 0.
\ea

Writing 
\be
	G = \sum_{n=1}^\infty {\cal G}_n
\ee
with ${\cal G}_n$ proportional to $\theta^n$ or its spatial derivatives; for the zeroth order in theta, we obtain
\ba
	\int d^2y \; && \left \{ \rho_0 \Omega(\vy) \delta(\vx -\vy) + \rho_x^i \dirac \( -f_i(\vy) + \frac 1m \ep_{ij}  {\partial^j_y} f_0 (\vy) - \right. \right. \nonumber \\
&& \left. \left. - \frac {3ie}{2m} \ep_{ij}[f^j,f_0](\vy) \) - kb \frac{\delta \cg_1(\vy)}{\delta \theta(\vx)} \right \} = 0,
\ea
whose solution is
\be
	\cg_1 = \frac \theta{kb} \(-\rho^\mu f_\mu + \frac 1m \ep_{\mu i \lambda} \; \rho^\mu \partial^i f^\lambda - \frac{3ie}{4m}\ep_{\mu \nu \lambda} \; \rho^\mu[f^\nu,f^\lambda] \).
\ee

For $\cg_2$ we have
\be
	\int d^2y \left \{ \rho^\mu_x \frac {\delta \cg_1(\vy)}{\delta f^\mu(\vx)} -kb \frac {\delta \cg_2(\vy)}{\delta \theta(\vx)} \right \}= 0,
\ee
hence
\be
	\cg_2 = -\frac 1{2k^2b^2} \( \rho^\mu \theta \rho_\mu \theta + \frac {3ie}{2m} \ep_{\mu \nu \lambda} \; f^\mu [\rho^\nu \theta, \rho^\lambda \theta] \).
\ee

Finally, $\cg_3$ is given by
\be
	\int d^2y \left \{ \rho_x^\mu \frac {\delta \cg_2(\vy)}{\delta f^\mu(\vx)} -kb \frac {\delta \cg_3(\vy)}{\delta \theta(\vx)} \right \} = 0.
\ee
Although the above derivatives of $\cg_2$ do not vanish, when they are contracted with vector $(\rho^\mu)$ the result is zero; hence $\cg_3 = 0$. This result implies that
\be
	\cg_n = 0 \;\;\;\; \forall n \ge 3.
\ee

Now the function $G$ is known. Consequently $G_0$ can be determined as well:
\be
	G_0 = - \frac 1{kb} \( \rho_0 \theta + \frac{3ie}{2m} \ep_{ij} \; [\theta, \rho^i f^j] + \frac{3ie}{4kbm} \ep_{ij} [\rho^i \theta, \rho^j \theta] \).
\ee

\subsection{The Function $\Psi$}
We have assumed that $\tilde \nu_\gamma$ and $\tilde \nu_\theta$ are zero-modes of matrix $\tilde f^\1$, now this condition will be used to find the function $\Psi$. Contracting $\tilde \nu_\theta$ with $\tilde f^\1$ and demanding this to be null, the following nontrivial equations emerge:
\be
	kb^2 = 1,
\ee
\ba
	\int d^2x &&\left \{ \frac 1m \ep_{ij} \rho^i_x \dirac + \frac 1b \frac{\delta \Psi (\vx)}{\delta f^j (\vy)} - \frac 1m \ep_{ij}  {\partial^i_x} \dirac + \frac {3ie}{2m} \ep_{ij}[f^i(\vx),\dirac] + \right.\nonumber \\ 
\label{psii}
&& \left. + \frac {3bie}{2m} \ep_{ij}[\theta (\vx), \rho_x^i \dirac] \right \} = 0
\ea
and 
\be
	\label{psitheta}
	\int d^2x \left \{ \rho_x^i \frac{\delta \Psi (\vy)}{\delta f^i (\vx)} - \frac 1b \Theta_{xy} + \frac bm \ep_{ij} \left ( \rho_x^i {\partial^j_x} \dirac + \frac{3ie}{2} [\dirac, \rho^i_x f^j(\vx)] \right ) \right \} = 0,
\ee
where the first equation have been used to achieve the last two. These, together with (\ref{psi0}), determines $\Psi$. The vector $\tilde \nu_\gamma$ does not generate any new condition to $\Psi$.

{}From equations (\ref{psi0}) and (\ref{psii}), except for a function dependent only on $\theta$, $\Psi$ can be found. Equation (\ref{psitheta}), as it can be checked, is redundant. So, there is some arbitrariness left on $\Psi$: if $\Psi$ is a solution of above equations, the same is true for $\Psi(f_0,f^i,\theta) + f(\theta)$, where $f(\theta)$ is any function of $\theta$. However, this arbitrariness has no importance, since such function only contributes with a surface term for the action (\ref{l1}).

Therefore, within our purpose, $\Psi$ is
\be
	\Psi = \frac bm \ep_{ij} \(\partial^i - \rho^i \)f^j - \frac {3bie}{2m} \ep_{ij} \left (\frac 12 [f^i,f^j] + b[\theta, \rho^i f^j] \right ) - bf_0
\ee 

\subsection{The Embedded Theory}

Gathering all was done, the Lagrangian $\tilde \cl^\1$ is
\ba
	\tilde \cl^\1 && = \frac 12 f^\mu f_\mu - \frac 1{4m} \ep^{\mu \nu \lambda} f_\mu F_{\nu \lambda} + \dot \lambda \tilde \Omega + \frac bm \ep_{ij} \( \partial^i f^j - \frac {3ie}2 \left ( \frac 12 [f^i,f^j] + b[\theta, \rho^i f^j] \right ) - bf_0 + \gamma \right )\dot \theta - \nonumber \\
	\label{finalL}
	&& - \frac bm \ep_{i j} \rho_0 f^i \partial^j \theta + \frac {b^2}2 \rho^\mu \theta \rho_\mu \theta - b\theta \rho^\mu f_\mu + \frac {3ibe}{4m} \ep_{\mu \nu \lambda} \( b f^\mu [\rho^\nu \theta, \rho^\lambda \theta] - \theta \rho^\mu [f^\nu,f^\lambda] \) - \frac 1{2b^2} \gamma \gamma.
\ea

Accordingly with the symplectic method \cite{mon}, the above Lagrangian has gauge symmetry with two independent generators (both related with zero-modes $\tilde \nu_\theta$ and $\tilde \nu_\gamma$), which are
\be
\label{gg}
\matrix{ \delta_{\vep_\theta}f_0 = \vep_\theta \rho_0 & & \delta_{\vep_\gamma} f_0 = \vep_\gamma \\[0.1in]\cr
\delta_{\vep_\theta}f^i = - \rho^i \vep_\theta & \hspace{0.2in} & \delta_{\vep_\gamma} f^i = 0 \\[0.1in]\cr
\delta_{\vep_\theta}\theta = - \vep_\theta \frac 1b &\hspace{0.2in} & \delta_{\vep_\gamma} \theta = 0 \\[0.1in] \cr
\delta_{\vep_\theta}\gamma = 0 &\hspace{0.2in} & \delta_{\vep_\gamma} \gamma = \vep_\gamma b \\[0.1in] \cr
\delta_{\vep_\theta}\lambda = \vep_\theta & \hspace{0.2in}& \delta_{\vep_\gamma}\lambda = 0}
\ee
The infinitesimal parameters are $\vep_\theta (\vx)$ and $\vep_\gamma (\vx)$. One can check that \footnote{Actually, $\delta_{\vep_\gamma} \tilde \cl^\1 = \tilde \Omega$, but we know from equations of motion that $\tilde \Omega = 0$.} $\delta_{\vep_\theta} \tilde \cl^\1 = \delta_{\vep_\gamma} \tilde \cl^\1 = 0$.

At this point we have already achieved the embedded version of the noncommutative self-dual model. Our next step is to rewrite the Lagrangian and its gauge generators in another form, one which will allow us obtain a St\"{u}ckelberg-like Lagrangian. Hence, we are looking for a symmetry with
\be
	\delta_\vep f^\mu = \vep \partial^\mu, \hspace{1in} \delta_\vep \theta = - \frac 1b \vep.
\ee
Comparing this with (\ref{gg}), it is not hard to guess we shall eliminate $\gamma$ through $\gamma = b^2 \dot \theta$ and choose
\be
	(\rho^\mu) = \pmatrix{ 0 & \partial^i},
\ee
following the procedure explained in Section II. Thus, the desired Lagrangian is found:
\ba
	\cl_S =&& \frac 12 \(f_\mu - b\partial_\mu \theta\) \(f^\mu - b\partial^\mu \theta\) -\nonumber \\
&&- \frac 1{4m} \ep^{\mu \nu \lambda} \(f_\mu - b \partial_\mu \theta \) \(\partial_\nu f_\lambda - \partial_\lambda f_\nu - ie[f_\nu - b\partial_\nu \theta, f_\lambda - b \partial_\lambda \theta]\).
\ea

\section{Conclusion}

In this work we achieved a dual version, with gauge symmetry, of constrained and unconstrained field theory models, including the case of the noncommutative manifolds.  We have developed the methodology in great generality and applied it to a diversity of interesting models presenting different physical contents such as second-class constraints in the Proca model, an unconstrained fluid model and a noncommutative self-dual model, which was latter particularized into a St\"{u}ckelberg-like version. It is important to emphasize two remarkable features of the used method: the easiness of handling the noncommutative part of the theory and the possibility to choose, through the infinitesimal gauge generators, which gauge theory will be built in. Regarding the first feature we remark that other approaches \cite{ghosh} make use of the Seiberg-Witten map beforehand to set up a commutative version in order to handle the embedding. This may limit the range of applicability to certain powers of the parameter of the Moyal product.  The duality treated in this work, however, holds to any power of the parameter of the Moyal product, since no restriction was necessary. 

The commutative part of this paper considered constrained and unconstrained models to illustrate the full power and generality of this technique.  Other embedding approaches are usually restricted to constrained models since they use the idea of constraint conversion to produce the gauge embedding. Our approach, on the other hand, only deals with the symplectic structure of the theory and does not depend on the previous existence of a constrained structure to produce the gauge structure.  This flexibility allowed us to approach the commutative and the noncommutative indistinctly which input great generality to the methodology.

\noindent ACKNOWLEDGMENTS: This work is partially supported by CNPq, CAPES,
FAPERJ and FUJB, Brazilian Research Agencies.  DCR thanks CNPq for the Doctoral fellowship.

\end{document}